\documentclass{article}

\usepackage{arxiv}

\usepackage[utf8]{inputenc} 
\usepackage[T1]{fontenc}    
\usepackage{hyperref}       
\usepackage{url}            
\usepackage{booktabs}       
\usepackage{amsfonts}       
\usepackage{nicefrac}       
\usepackage{microtype}      
\usepackage{graphicx}
\usepackage{tabularx}

\title{A Survey and Comparative Study on Multi-Cloud Architectures: Emerging Issues and Challenges for Cloud Federation}

\author{ \href{https://orcid.org/0000-0002-9689-6387}{\includegraphics[scale=0.06]{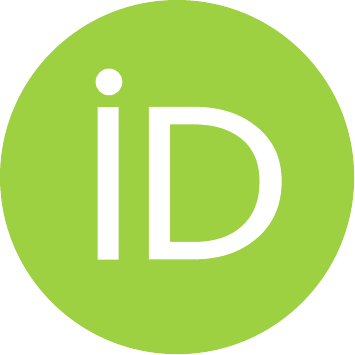}\hspace{1mm}Deepika Saxena}\thanks{The authors would like to thank National Institute of Technology Kurukshetra, India for financially supporting this research work.} \\
	Department of Computer Applications\\
    National Institute of Technology\\
	Kurukshetra, India \\
	\texttt{deepika\_6180096@nitkkr.ac.in} \\
	\And
	\href{https://orcid.org/0000-0001-9305-1269}{\includegraphics[scale=0.06]{orcid.pdf}\hspace{1mm}Rishabh~Gupta}\\
		Department of Computer Applications\\
	National Institute of Technology\\
	Kurukshetra, India \\
	\texttt{rishabh\_6180047@nitkkr.ac.in} \\
	
	\And
	\href{https://orcid.org/0000-0002-8053-5050}{\includegraphics[scale=0.06]{orcid.pdf}\hspace{1mm}Ashutosh Kumar Singh} \\
		Department of Computer Applications\\
      National Institute of Technology\\
	Kurukshetra, India \\
	\texttt{ashutosh@nitkkr.ac.in} \\

}

\renewcommand{\shorttitle}{Survey on Multi-Cloud Architectures}

\hypersetup{
pdftitle={A template for the arxiv style},
pdfsubject={q-bio.NC, q-bio.QM},
pdfauthor={David S.~Hippocampus, Elias D.~Striatum},
pdfkeywords={First keyword, Second keyword, More},
}

\begin{document}
\maketitle

\begin{abstract}
Multi-cloud concept has broaden the world of cloud computing and has become a buzzword today. The word “Multi-cloud” envisions utilization of  services from multiple heterogeneous cloud providers via a single architecture at customer (e.g. organization) premises. Though cloud computing has many issues and offers open research challenges, still the academics and industrial research has paved a pathway for multi-cloud environment. The concept of multi-cloud is in maturing phase, and many research projects are in progress to provide a multi-cloud architecture which is successfully enabled in all the respects (like easy configuration, security, management etc.). In this paper, concepts, challenges, requirement  and future directions for multi-cloud environment are discussed. A survey of existing approaches and solutions provided by different multi-cloud architectures is entailed along with analysis of the pros and cons of different architectures while comparing the same.

\end{abstract}

\keywords{cloud broker\and cloud federation\and inter-operable \and multi-cloud.}

\section{Introduction}
Cloud computation world is transforming in anticipation and is rightly contributing in successful technical progress of almost every organization across the world \cite{saxena2021proactive}, \cite{saxenaa2020communication}, \cite{saxena2018abstract}, \cite{saxena2021securevmp}. From economic point of view, cloud computing represents a business model for renting software and hardware resources \cite{singh2021cryptography}, \cite{singh2021quantum}, \cite{singh2019secure}, \cite{saxena2014review}. Multi-Cloud has extremely attracted the corporate world. EMA surveyed 260 enterprises and majority of the respondents i.e., 61\% reported using two or more public cloud provider. All of us are familiar with the fact that the amount of useful information produced by an organization today in one day is approximately equal to the amount of information produced in twenty days earlier a couple of decades back \cite{saxena2015ewsa}, \cite{saxena2016dynamic}, \cite{saxena2020auto}, \cite{saxena2020security}, \cite{saxena2021energy}. Single cloud framework is not suitable and has become insufficient to serve the ever growing requirement \cite{kumar2018long}, \cite{kumar2018workload}, \cite{kumar2020biphase}, \cite{kumar2021resource}. Seeing the current trend of computational growth in
organizations, it is quiet obvious that future is multi-cloud.
\subsection{Multi-cloud concept}
With the advent of cloud, small and big organizations all are progressing without need to concern about the storage and maintenance of their business data. They are not required to spend in billions for the same. All the responsibility is envisaged upon the cloud service providers (CSPs) \cite{saxena2021op}, \cite{saxena2021osc}, \cite{saxena2015highly}. So, cloud computing has become the backbone of modern business world. Organizations contacts various cloud service providers and consumes the services by signing Service-Layer Agreement SLA document \cite{chhabra2020secure}, \cite{gupta2019confidentiality}, \cite{gupta2021data}, \cite{varshney2021machine}, \cite{chauhan2020survey}. A CSP contacts various resource providers at datacenters in order to satisfy the demands of the customer \cite{saxenaglobal}, \cite{kaur2017comparative}, \cite{chhabra2018probabilistic}, \cite{patel2021review}. Usually, it is said that cloud computing provides infinite resources and elastic services \cite{saxena2021workload}, \cite{kumar2020ensemble}. Practically, we know that resources cannot be infinite at any datacenter. To raise the elasticity or capacity of cloud service providers and fulfill the ever growing demand of services, resources from different resource providers need to collaborate, inter-communicate and work in cooperation and coordination. So, collaboration of various cloud service providers gives root to the concept of Multi-Cloud.
Multi-cloud simply means an enterprise can take services from more than one cloud service provider through a common interface or a single API. In multi-cloud environment, multiple cloud providers aggregate in a single pool to provide three basic services of cloud: computation, storage and networking. Computation service allows customers to instantiate their own computational nodes irrespective of their physical server node at different datacenters. Storage service enables users to store infinite data as per they desire without vendor- lock-in. They can store public data at public cloud and confidential data at private cloud

\subsection{Vendor lock-in}
In Single cloud environment, security of user data is completely in the hands of the cloud provider and he leverages entire control of user data. This situation represents “vendor lock-in” drives user data towards potential security risks. To resolve this issue, multi- cloud approach provides appropriate solution.
Networking provides communication between computational and storage nodes despite of the networking infrastructure hosted upon physical connectivity between actual physical servers.
Multi-cloud allows user to organize, transfer, synchronize and manage sharing of files between cloud storage services like  Dropbox, Google Drive, Copy, OneDrive, FTP, WebDav, MEGA, etc. 26 cloud drives supported so far \cite{saxena2015highly}.
\subsection{Challenges of Multi-Cloud }
\begin{itemize}
    \item 	API’s differ: Different sets of resources are there, different formats and encodings, and several simultaneous versions of single cloud exist.
    \item  	Abstractions differ: Network architecture differ- VLANs, topology differ, addressing grounds differ.
    \item 	Difference in storage architecture- local/attachable, disks and backups.
    \item 	Hypervisor and machine configuration differ etc.
    \item 	Mismatch of cyber-laws of various interconnected clouds depending on their geographical location physically.
\item 	Interoperability of different cloud providers always a major issue for multi-cloud set-up.
\item  	Difficult to tackle the security challenges related to multi- provider environment.
\item 	Different cloud provider follows their own management scheme for load balancing and monetization features.
\end{itemize}
\textbf{Interoperability:} The biggest challenge in cloud computing is the diversity of resources, management and rules and regulations of cloud providers, diverse SLAs, differences in security characteristics at various CSPs etc. Interoperability means the ability of diverse system to work in cooperation and serve the common output. Interoperability of multi-cloud has several heterogeneous dimensions: 
\begin{itemize}
    \item \textbf{Vertical interoperability} Along the cloud stack, interoperability increases as we move from IaaS to SaaS. To deal with such heterogeneity, standardize platform and mapping interface layers are required.
    \item \textbf{Horizontal interoperability} The inter-communication issues lying between various service providers, technically, physically, logically and legally etc. while contributing for multi-cloud.

\end{itemize}
Across the world, thousands of data centers are geographically distributed which vary in cost of electrical consumption. At various data centers, there exist  heterogeneity in resource usage there logical and physical infrastructure.
 Each different CSP follows his own SLA policies, management rules, security features, cloud service cost etc. In order to deal with the above challenges, various inter-cloud or federated cloud approaches were proposed. Many projects are in progress to set-up multi-cloud architecture. In this paper, number of multi-cloud or federated cloud or inter-cloud approaches are surveyed.
\subsection{Organization} The remainder of this paper is as follows: Section1 introduces concept of multi-cloud and challenges of multi-cloud. Section 2 gives detailed architectures of various cloud federation available till now. Section 3 provides critical analysis on various architectures discussed in this paper. Section 4 concludes the paper and highlights the future scope.

\section{ARCHITECTURAL VIEWS OF MULTI-CLOUD}
\label{sec:headings}
\subsection{CHARM}
In the paper \cite{zhang2015charm}, (Quanlu Zhang et.al,2015) have  presented “CHARM: A Cost-Efficient Multi-Cloud Data Hosting Scheme with High Availability”. Figure 1 shows the basic principle of multi-cloud data hosting. Here the basic principle of data hosting on multi-cloud is to distribute data across multiple clouds to gain redundancy and prevent vendor-lock-in. The key component in this model is “Proxy” which redirects the request from client application and coordinate data distribution among multiple clouds. Different clouds exhibit huge difference in terms of their services like pricing policies and work performance. For instance, Google cloud platform charges more for bandwidth consumption but Amazon S3 charges more for storage space and Rackspace provides all web operations free via series of REST ful APIs. In this paper, in order to reduce monetary cost and guaranteed availability, two redundancy schemes are combined- replication and erasure coding. The advantage of the CHARM multi- cloud model is that it provides guidance to the customers to distribute their data on multiple clouds in cost effective manner. CHARM makes fine-grained decisions about which storage mode to use and which clouds to place data in.
\begin{figure}[!htbp]
    \centering
    \includegraphics{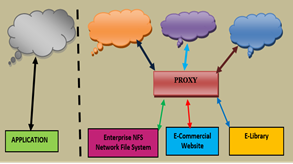}
    \caption{Basic principle of multi-cloud data hosting}
    \label{motivation}
\end{figure}
\subsection{	OPTIMIS Cloud Federation }
OPTIMIS cloud federation \cite{ferrer2012optimis} have presented toolkit regarding cloud service and cloud provider is presented. This toolkit can optimize the whole service life cycle of cloud including service construction, deployment, and operation on basis of aspects such as trust, risk, eco-efficiency and cost. Several architecture of cloud are also addressed in this paper. More significantly, two architecture of cloud are discussed here.
\begin{itemize}
    \item \textit{	Federated Cloud}: In this architecture, service provider (SP) takes services from a single infrastructure provider (IP) , but, one infrastructure Provider(IP) can sub-contract (or can lease some services) with another IP , in order to raise the capacity, elasticity and fault tolerance of cloud service.
    
    \item \textit{Multi-provider hosting}: In this architecture, Service Provider (SP), directly negotiates with more than one infrastructure providers, to deploy services, monitors their operation and migrates services from misbehaving IPs to maintain trust relationship.

\end{itemize}
\subsection{Multi-Cloud Brokering Architecture}
In this type of architecture, a third party broker aggregates services from multiple infrastructure providers (IPs) and delivers these services to the actual service provider. The broker act as SP for multiple IPs and it behave as IP for the SP. This model simplifies the complexities of taking services from several different IPs. Here, the broker provides single entry point to many IPs and more fruitful is that the broker can get bulk of discount from IPs from economic point of view. Moreover, in terms of business policies have to make long term relation with the broker for excellent growth of his business. Hence, risk and trust prediction becomes easier with broker in between IPs and SP.

\subsection{RESERVOIR MODEL}
\begin{figure} [!htbp]
    \centering
    \includegraphics{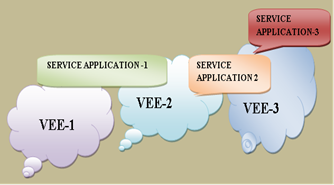}
    \caption{Conceptual view of Reservoir Model of cloud federation}
    \label{fig 5}
\end{figure}

Rochwerger et al., 2009 \cite{galis2009reservoir} have introduced Reservoir model for federated cloud computing. According to this architecture, different infrastructure provider (IP) can aggregate to provide flexible and scalable cloud computing services as shown in figure \ref{fig 5}. Here, IPs own the computational resources, network resources and storage resources in an infinite pool. Service providers are the agents who offers computational and storage services to the customers with help of IPs resources at back end. In Reservoir model, IPs and SPs are clearly separated. IPs operates reservoir sites that actually owns and manages resources. The reservoir sites collaborates to form a Reservoir Cloud. Virtualization layer is set up for separation of different Virtual Execution Environment (VEEs) and the resource usage optimization at each reservoir site. Service applications of customers are deployed on reservoir cloud using service manifest, on one or more VEE as per the requirement.
Service manifest will give information regarding capacity of physical infrastructure such as the number of VEEs that can be hosted on a CPU, structure of service application in terms of software stack (OS, middleware, application, data and configuration). Service manifest also indicates Key Performance Indicators (KPI). KPI measures load parameters, response time, throughput, number of active sessions and resource allocations i.e. memory, bandwidth, number of VEEs. To resolve interoperability issues, this model provides VEE management interface that supports VEEM-to-VEEM communication. The broad limitation of this model is that it has not mentioned any security issues handling while collaborating various IPs in reservoir cloud. The practical implementation of this reservoir model is not given.
\subsection{Federated Cloud Management (FCM) }
\begin{figure} [!htbp]
    \centering
    \includegraphics[width=.5\linewidth, ]{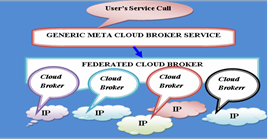}
    \caption{FCM Architecture}
    \label{fig 6}
\end{figure}

 In FCM \cite{marosi2011fcm}, several Infrastructure Providers exclusively dedicate their computing infrastructure to a single cloud broker. Figure \ref{fig 6} shows FCM architecture where, various cloud brokers collaborates to provide federated cloud environment. In FCM, interoperability is resolved by enabling two levels of brokering.
\begin{itemize}
    \item User service call is submitted to Generic Meta –Broker Service (GMBS).
    \item This call is then forwarded to selected Federated Cloud Broker, which further selects VM from underlying infrastructure provider for actual execution of service.
FCM incorporates high level brokering for cloud federation and on demand automated service deployments. FCM Repository maintains all the required information about virtual appliances at cloud broker level.
\end{itemize}
\subsection{Distributed Management Task Force (DMTF) Architecture}
 Cloud, \& Incubator \cite{clouds2009white} have presented Distributed Management Task Force (DMTF) Open Cloud standards Incubator models for enabling interoperability in multi-cloud environment. According to DMTF, three different models for supporting federated cloud scenario have been identified. Datacenter includes Service Providers (SP) and infrastructure provider (IP). The three models differs in the working style of datacenters as follows:
 \begin{itemize}
     \item Firstly, at the datacenter, service provider can act  as  cloud broker or an agent to federate underlying infrastructure provider resources and make the services available to the overlying customers as shown in figure \ref{fig 7}.
     \begin{figure} [!htbp]
    \centering
    \includegraphics{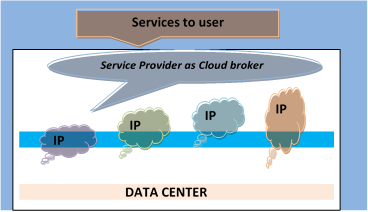}
    \caption{First case for cloud federation according to DMTF approach}
    \label{fig 7}
\end{figure}
\item 2.	In second case, multi-cloud model is presented, where datacenter can request services from various infrastructure or resource providers, thus they support different SLA’s parameters. Figure \ref{fig 8} shows this scenario.
\begin{figure} [!htbp]
    \centering
    \includegraphics{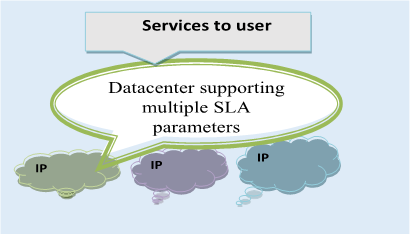}
    \caption{Second case for cloud federation according to DMTF approach}
    \label{fig 8}
\end{figure}
\item 	In third case,  if a datacenter is served by one SP and hosted by one IP overflows the available computing capacity limit, then another infrastructure provider can contribute extra computing capacity for the former IP. But, Service Provider is unaware of this resource provisioning.

 \end{itemize}
\subsection{Contrail Architecture}
\begin{figure} [!htbp]
    \centering
    \includegraphics{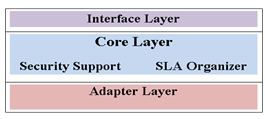}
    \caption{Contrail Cloud Federation}
    \label{fig 10}
\end{figure}

Carlini et al. \cite{carlini2011cloud} introduced Open Contrail Architecture for cloud federation. The goal of Open Contrail project is to minimize the burden on user and increase efficiency of the cloud platform, by performing horizontal ( resolving interoperability issues in cloud federation) and vertical (providing cloud federation facility at cloud stack level i.e. IaaS and PaaS services) integration. This architecture utilizes the resources of various cloud resource providers, regardless of the heterogeneity of underlying infrastructure. Complete open source approach is adopted here. It also supports user authentication, integration of SLA management and application deployment.
Contrail architecture has three layer to support federation of clouds. Figure \ref{fig 10} shows Open Contrail Federation:
\begin{itemize}
    \item Interface Layer: This is top most layer, which provides interaction between user and contrail federation. It consists of REST API, http and Command Line Interface.
    \item Core Layer: This is middle layer, which contains various modules to provide functional requirements like application life-cycle management and non-functional like security requirements.
    \item Adapter Layer: This is bottom layer of the architecture, that contains modules to retrieve information and operate on different cloud resource providers. It copes with the heterogeneity of underlying infrastructure.

\end{itemize}

\subsection{Combinatorial  Auction based Collaborative Cloud}
\begin{figure} [!htbp]
    \centering
    \includegraphics[width=.24\linewidth]{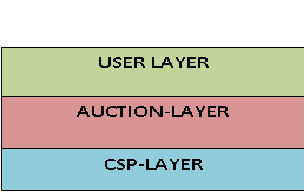}
    \caption{Auction-Based Collaborative Cloud}
    \label{fig 11}
\end{figure}

Zhang, Li and Zhu, 2015 \cite{zhang2015combinatorial} have presented a Combinatorial Auction- Based Collaborative Cloud Services Platform in the form of model for cloud collaboration considering cost of communication and negotiation among clouds under collaboration. This market model for cloud collaboration includes three distinct layers as shown in figure
\ref{fig 11}. First, user layer receives requests from end-user. Second, Auction-layer matches requests with cloud services provided  by CSP. Third, CSP-Layer forms coalition to improve serving ability to satisfy demands of the user.
For coalition formation and to find suitable partner for collaboration, Breadth Traversal Algorithm (BTA) and Revised Ant Colony Algorithm (RACA) are introduced. Here CSPs dynamically collaborate with one another and submit their group’s bid as a single bid to the auction-layer. Whenever a request is generated by the user, at user-layer, BTA and RACA algorithm executes at auction-layer, for the selection of CSP to enable dynamic collaboration of cloud to fulfill the request of the user. The results presented with this model are not tested with real-time data so it is somewhat unacceptable model.
\subsection{Layered Multi-Cloud  Architecture}
\begin{figure} [!htbp]
    \centering
    \includegraphics{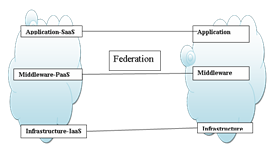}
    \caption{Cloud federation layered model}
    \label{fig 12}
\end{figure}

Villegas et al. \cite{villegas2012cloud}  have shown a layered multi-cloud architecture by creating federation at each service layer as shown in figure \ref{fig 12}. At each layer, broker is present, who follows some pre-defined collaboration terms for successful inter-cloud federation. They have added federation of cloud model by showing how it works in delivering Weather Research Forecasting (WRF). They have given conceptual view of decoupling SaaS, PaaS and IaaS clouds. For such federation, well defined policies and protocols must be defined. PaaS layer act as middleware and act as a bridge for vertical integration. It is a bridge between application and elastic infrastructure resource management.

\subsection{mOSAIC Multi-Cloud Architecture}
\begin{figure} [!htbp]
    \centering
    \includegraphics{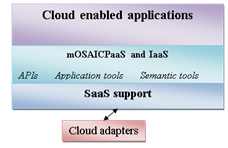}
    \caption{mOSAI Cloud model}

    \label{fig 13}
\end{figure}

Petcu et al. \cite{petcu2013experiences} have introduced  Open-source API and Platform for Multiple Clouds (mOSAIC). This project provides solution for application portability and interoperability across multiple clouds. It offers a middleware for deployment of multi-cloud application. mOSAIC also handles the problem of changing of cloud vendors. Here, application instance runs in mOSAIC component getting access through dedicated APIs. Communication among its components takes place through message queues. E.g. AMQP.
Cloud Agency of mOSAIC is a service for deployment of cloud applications. Here Cloud Agency handles interoperability problem by introducing a uniform interface for accessing multiple clouds. As shown in figure \ref{fig 13}, mOSAIC model has provision for :\\
-application portability between clouds via set of APIs.\\
-Support dynamic negotiations with multiple cloud providers using brokering system.\\
-user-centric SLA.

\subsection{	EGI FEDERATED CLOUD}
EGI Federated cloud \cite{fernandez2015egi} integrates public, private or hybrid cloud in the form of multi-national cloud system. Each resource center operates cloud management framework (CMF) according to its own preferences and then joins Federation and Collaboration services . User can access all services of cloud federation with a single identity.
\subsection{	CONTAINER TECHNOLOGY}
Containers \cite{trofimov2017use} have replaced traditional hypervisor based Virtual machines. All most every top company like Microsoft, Facebook, Google are using Docker Swarm or Kubernetes technology. Basically, containers provide easy to deploy and secure orchestration of specific self-contained virtual infrastructure. Cloud containers can virtualize single application environment. Containers create an isolation boundary at application level rather than at server level
.Isolation ensures that if anything goes wrong in a container, then it may not effect another container deployed on same VM or server. In absence of container, VM requires complete operating system to be installed, but with container technology neither requires complete installation of operating system nor even they requires complete virtual copy of host server’s hardware resources.
Containers only requires software , libraries and basics of an operating system and cloud containers can be very easily copied and deployed on another server very easily.The only limitation of containers is their management. Containers splits the virtualization into smaller chunks, that creates difficulty in management.[17].
\subsection{	SPINNAKER}
Spinnaker \cite{furber2014spinnaker} is an open source ,multi-cloud delivery platform. It provides two important features : 1. Cluster management 2. Deployment management.\\
\textit{Cluster management}: User can manage resources in the cloud cluster which is a logical grouping of servers in spinnaker. The logical grouping is based on user application. Server groups defines the deployable artifacts i.e. VM image, Docker image, source location.\\
\textit{Deployment management}: The key feature of spinnaker are pipelines. Pipelines consists of sequence of actions (like deploy, resize, disable, manual judgement etc.) known as stages.
\subsection{CLOUD FOUNDRY}
Cloud Foundry \cite{bernstein2014cloud} provides a standardized platform to the applications of customers by decoupling the application from its infrastructure. To enable this, CF offers two types of VMs: the component VMs as platform infrastructure and the host VMs that host applications for outside world.
Hence, the organizations can make a business decision on where to deploy workloads i.e. on premise, in managed infrastructure or in public cloud.
\subsection{SUPER CLOUD PROJECT}
\begin{figure} [!htbp]
    \centering
    \includegraphics{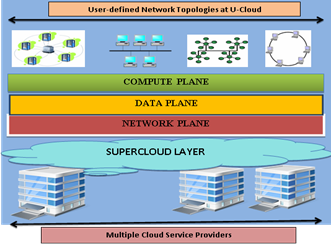}
    \caption{SUPERCLOUD Architecture}

    \label{fig 14}
\end{figure}

This project (Lacoste et al. \cite{lacoste2016user}) has divided multi-cloud architecture into three layers: data, compute and network layer. This work focus on user-centric vision of cloud i.e. virtual private cloud (VPC) just like VPN, inspite of usual provider-centric cloud. To deal with interoperability issues, a resource distribution layer i.e. SUPERCLOUD has been introduced. VPC are set up in the form of U-Clouds consisting of data units, compute unit and network units. These U-Clouds are VPC set upon top of various cloud service providers. For security, they have introduced shared data encryption and middleware-based encryption and decryption of data. Figure \ref{fig 14}, shows deployment architecture of SUPERCLOUD project work.
\begin{itemize}
    \item \textit{Compute Plane}: This is topmost plane which applies nested virtualization for abstracting computational resources of various CSPs. Here, user can run several virtual execution environment like, containers and virtual machines. This plane ensures proper execution and protection of each user VMs. These VMs at compute plane are deployed over user-hypervisor layer, which itself is set-upon provider’s hypervisor.
    \item \textit{Data Plane}: This plane manages all the storage management responsibility and provides data management APIs for every execution environment on the compute plane.In order to get access to the cloud’s storage, a user’s compute VM realizes data management VM as a proxy to get access to the cloud data.
    \item \textit{Network Plane}: The network management is a network virtualization platform based on Software-Defined Network (SDN).Here, the host VM sends request to the network management VM to get direct access to the storage data. This plane enables isolation between multiple tenants by providing abstraction for network topology and traffic routing.

\end{itemize}

\subsection{Utility-oriented Federation of Cloud Environment }
\begin{figure} [!htbp]
    \centering
    \includegraphics{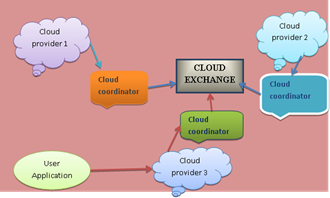}
    \caption{Utility based Federated Cloud Model}

    \label{fig 15}
\end{figure}

Buyya et al. \cite{buyya2010intercloud} have  proposed	inter-cloud framework which supports scaling of applications across multiple vendor clouds. In this inter-cloud model, as shown in figure \ref{fig 15}, every client in federated environment needs to contact cloud brokering service that can dynamically establish service contracts with cloud coordinators through trading functions exposed by cloud Exchange. 
Cloud Coordinators- They caters the deployment and management responsibility for domain specific enterprise clouds and their membership to overall federation driven by market-based trading and negotiation protocols. Also, sensor infrastructure will monitor the power consumption, heat dissipation, utilization of computational nodes in virtual cloud environment.
Cloud Broker finds out suitable Cloud service provider through cloud exchange and negotiates with cloud coordinators for resource allocation.
Cloud Exchange keeps information about cloud’s current usage cost and demand patterns. In the end, it is concluded that federated cloud service gives better performance than existing non-federated cloud approaches.\\

Table \ref{Table:workloadpredictionsummary} entails the summary and comparative analysis of aforementioned multi-cloud architectures.\\
 
\begin{table}[!htbp]
 \centering
 
  \caption{Comparative Summary of Multi-cloud Architectures}
 	\begin{tabular}{ |  p{2.5cm} |  p{2cm} |  p{3.5cm} | p{3.5cm} | p{3.5cm} |} 
 		\hline \textbf{Model, Author }& \textbf{Aim} & \textbf{Conceptual Approach}& \textbf{Strength}&\textbf{Weakness } \\ \hline \hline
 		CHARM
(Zhang et.al, 2015)
\cite{zhang2015charm} & Cost-efficient multi-cloud		data hosting	scheme with		high
availability
 &  “Proxy” redirects the request from client application	and
coordinate	data
distribution among multiple clouds. In order to reduce monetary cost and guaranteed availability, two	redundancy schemes are combined- replication   and erasure
coding
 & It provides guidance to the customers to distribute their data on multiple clouds in cost effective manner & Various issues regarding cloud computing are uncovered	like, Management of multi- cloud, load-balancing, security etc.\\ \hline
 OPTIMIS CLOUD FEDERATION
(Ferrer et al., 2012) \cite{ferrer2012optimis} & To provide toolkit to optimize whole service lifecycle as a cloud &Service Provider (SP), directly negotiates with more than one infrastructure providers, to deploy services.
a third party broker aggregates         services
from	multiple infrastructure providers (IPs) and delivers these services   to   the  actual
service provider
& This model simplifies the complexities of taking services from several different IPs. The broker provides single entry point to many IPs and more fruitful is that the broker can get bulk of discount from IPs from economic point of view
& Only	conceptual architecture of several multi-cloud or federated approach architecture are given , but exact practical implementation is not given. How to handle interoperability issues is not focused clearly \\ \hline

DISTRIBUTED
MANAGEMENT
TASK FORCE (DMTF)
ARCHITECTURE
(Paper et al., 2009) \cite{clouds2009white} & Open	Cloud
standards
Incubator models
for	enabling
interoperability in
multi-cloud
environment.
& service provider can act
as	cloud	broker	to
aggregate services from
infrastructure provider
and	make	them
available	to	the
customers.
Different	IPs	can
collaborate to provide
services to one service
provider.
 & They	have	conceptually
provided Incubator models
for handling interoperability
in multi-cloud environment.
 & Service	Provider	is
unaware of this resource
provisioning. So, lack of
reliability and trust.
Security is major concern
that is not included here.

\\ \hline
 RESERVOIR MODEL

(Rochwerger	et	al., 2009) \cite{galis2009reservoir} & Reservoir model for federated cloud computing & Different infrastructure provider (IP) can aggregate to provide flexible and scalable cloud	computing services. Virtualization layer is set up for separation of different Virtual		Execution Environment (VEEs) and the resource usage
optimization at each reservoir site & The resource usage optimization at each reservoir site.
Interoperability is handled by VEE management interface that supports VEEM-to-VEEM
communication & Security, access feature, management of various multi-cloud architecture are not discussed.The practical implementation of this reservoir model is not given \\ \hline

LAYERED	MULTI-
CLOUD
ARCHITECTURE
(Villegas et al., 2012)
\cite{villegas2012cloud}& Layered	multi-
cloud architecture
by	creating
federation at each
service layer .
 & In layered multi-cloud
architecture,	at	each
layer, broker is present,
who follows some pre-
defined	collaboration
terms	for	successful
inter-cloud federation.
 & They have given conceptual
view of decoupling SaaS,
PaaS and IaaS clouds. They
have identified, that for
such	federation,	well
defined	policies	and
protocols must be defined.
 & This	architecture	lacks
management	and
authorized	access
handling.It is not clear
exactly how PaaS can act
as a middleware.
\\ \hline
 \end{tabular}
 	   \label{Table:workloadpredictionsummary}
 	\end{table} 
 	
 \begin{table}[!htbp]
  \renewcommand\thetable{1}
 \centering
  \caption{Comparative Summary continued}
  
 	\begin{tabular}{ |  p{2.5cm} |  p{2cm} |  p{3.5cm} | p{3.5cm} | p{3.5cm} |} 
 		\hline \textbf{Model, Author }& \textbf{Aim} & \textbf{Conceptual Approach}& \textbf{Strength}&\textbf{Weakness } \\ \hline \hline 
CONTRAIL
ARCHITECTURE
(Carlini et al., 2012) \cite{carlini2011cloud} & To minimize the
burden on user and
increase efficiency
of	the	cloud
platform,	by
performing
horizontal	(resolving
interoperability
issues	in	cloud
federation)	and
vertical (providing
cloud	federation
facility at cloud
stack level i.e. IaaS
and PaaS services)
integration. & This	architecture
utilizes the resources of
various cloud resource
providers, regardless of
the	heterogeneity of
underlying
infrastructure. Complete
open source approach is
adopted here.
 & It handles interoperability
issues by using open source
environment completely.
It	also	supports	user
authentication,	integration
of SLA management and
application deployment.

 & Though this model is quiet
efficient, still there is s
limitation that how can we
sure	that	each	cloud
resource	provider	is
trustworthy regarding his
deliverables.
 \\ \hline

COMBINATORIAL
AUCTION-BASED
COLLABORATIVE
CLOUD
(Zhang et al., 2015)
\cite{zhang2015combinatorial} & Their aim is to
provide model for
cloud collaboration
considering cost of
communication
and	negotiation
among	clouds
under
collaboration.

 & First, user layer receives
requests from end-user.
Second, Auction-layer
matches requests with
cloud services provided
by CSP. Third, CSP-
Layer forms coalition to
improve serving ability
to satisfy demands of
the user. For coalition
formation and to find
suitable	partner	for
collaboration
 & This is a market-model for
multiple	cloud
collaboration, which selects
the suitable CSP according
to demand of the user.
& The results presented with
this model are not tested
with real-time data so it is
somewhat	unacceptable
model. \\ \hline

mOSAIC	MULTI-
CLOUD
ARCHITECTURE
(Petcu et al., 2013)
\cite{petcu2013experiences} & 
This	project
provides solution
for	application
portability	and
interoperability
across	multiple
clouds.

& 
mOSAIC architecture is
built from
loosely	coupled
components	that
enhance the chances for
the	open-source
software prototypes to
be adopted.
& 
mOSAIC	handles	the
problem of changing of
cloud vendors.
It has application portability
between clouds via set of
APIs.
-Support	for	dynamic
negotiations with multiple
cloud	providers	using
brokering system.
& 
This model lacks security
management.
Event-driven
programming	styleof
mOSAIC application is
complex.
 \\ \hline
 EGI	FEDERATED CLOUD \cite{fernandez2015egi} & EGI	Federated
cloud integrates public, private or hybrid cloud in the form of multi- national cloud system.
 & Each resource center operates		cloud management framework (CMF) according to its own preferences and then joins Federation and	Collaboration
services. & User can access all services of cloud federation with a single identity. & This federated cloud model lacks trust and unauthorized access issues and other security features are not considered. \\ \hline

 \end{tabular}
 	   \label{Table:workloadpredictionsummary}
 	\end{table} 
 \begin{table}[!htbp]
  \renewcommand\thetable{1}
 \centering
  \caption{Comparative Summary continued}
  
 	\begin{tabular}{ |  p{2.5cm} |  p{2cm} |  p{3.5cm} | p{3.5cm} | p{3.5cm} |} 
 		\hline \textbf{Model, Author }& \textbf{Aim} & \textbf{Conceptual Approach}& \textbf{Strength}&\textbf{Weakness } \\ \hline \hline

FEDERATED CLOUD MANAGEMENT (FCM) ARCHITECTURE
(Marosi et al., 2011) \cite{marosi2011fcm} & FCM incorporates high	level
brokering	for cloud federation and    on    demand
automated service deployments & several	Infrastructure Providers		exclusively dedicate			their computing infrastructure to a single
cloud	broker,	which further selects VM from underlying infrastructure provider for actual execution of service & Conceptually, the model is simple and easy to understand & Interoperable	issues between various cloud brokers in practical scenario is not considered. Lack of management and security constraints \\ \hline

CONTAINER TECHNOLOGY \cite{trofimov2017use} & containers aims to provide easy to deploy and secure orchestration      of
specific	self- contained     virtual
infrastructure
 & Containers create an isolation boundary at application level rather than at server level & Containers only requires software , libraries and basics of an operating system and cloud containers can be very easily deployed on another server very
easily.
& Containers splits the virtualization into smaller chunks, that creates difficulty in management \\ \hline
SPINNAKER \cite{furber2014spinnaker} & Spinnaker is an open source ,multi- cloud delivery platform & It	provides	two important features:1.Cluster management 2.Deployment
management.
 & Multiple heterogeneous cloud services can be availed through single API. & Management and handling of multiple cloud clusters by connecting through pipelines is a complex. It lacks	security
management.
\\ \hline
Cloud Foundry \cite{bernstein2014cloud} & It aims to provide  a standardized platform to the applications        of
customers	by
decoupling	the
application from its infrastructure
 & CF offers two types of VMs: the component VMs as platform infrastructure and the host VMs that host applications for outside world. & The organizations  can easily make a business decision on where to deploy workloads i.e. on premise, in managed infrastructure & This model do not handles load balancing, security and data storage handling issues. \\ \hline
 
 SUPER
CLOUD \cite{lacoste2016user}& This work aims to provide	self-
managed, self- protecting virtual private clouds separately for each user.
& All the CSPs are at bottom layer,supercloudlayer handles	all		the interoperability	issues and provides complete ease of usage in the hands of user. & Supercloud is self-managed, self-protecting cloud of clouds,providing complete management of multi- clouds in the hands of user. & Particularly in our view, this project model leaves no concern untouched regarding multi-cloud set- up as such, also this project work is in progress till now. But this project favours users more than the cloud service
providers.
 \\ \hline
 UTILITY-ORIENTED FEDERATION	OF CLOUD COMPUTING ENVIRONMENTS \cite{buyya2010intercloud} & Inter-cloud framework which supports scaling of applications across multiple	vendor clouds. & Every client in federated environment needs to contact cloud brokering service that can dynamically establish service contracts with cloud coordinators
through	trading functions exposed by cloud Exchange & This model is efficient with respect to trust as cloud broker selects CSP by negotiating with cloud coordinators. & Though trust is given priority in this model, still there seems no concern for security management. \\ \hline
 \end{tabular}
 	   \label{Table:workloadpredictionsummary}
 	\end{table} 
After analyzing various models of multi-cloud set-up reveals that most of the work done in this regard( taking services from multi- cloud) handles only interoperability issues and lacks concern for trust and security management while collaborating. Also, much of the research is limited upto theoretical concept of joining multiple clouds together by having some Broker service between various infrastructure providers and service providers. Still in our opinion, SUPERCLOUD project model is most suitable among the various available multi-cloud architectures till now. This is because, it has covered almost every concerned feature for enabling successful set- up of multi-cloud.
 	\section{Conclusion and Future Directions}
 	Single-cloud services are not suitable to meet the ever growing and variety of demands of the organization. Hence, more effective way is to employ multiple clouds to serve an organization. Every data  cannot be treated with`equal status. As some data may be more confidential than other, and some data will be completely public for the advertisement. Hence according to the security requirement , cost and official management of different data, it can be placed on different appropriate clouds as per the suitability of the organization. But to take services from multiple clouds is quiet complex as you have to individually deal with multiple cloud service providers. So, its better if there would be some unified platform from where can access services of multiple clouds- Multi-cloud infrastructure. To set-up multi-cloud, their exists lots of complex issues- like interoperability, monetizing, security management issues etc. In order to deal with these issues many researchers from academic as well as industrial background have provided solutions in the form of federated cloud or multi-cloud architectures. In this paper, we have presented analytical review of various multi-cloud architectures and also we have given comparative analysis in the form of pros and cons of each architecture. Finally, we have concluded that in our view, SUPERCLOUD project model is at the top among all the existing multi-cloud architecture concepts. But this project is in progress and it is not fully implemented work.
\bibliographystyle{IEEEtran}
\bibliography{bibfile}


\end{document}